\documentclass[3p,times,procedia]{elsarticle}
\usepackage{nupha_ecrc}

\volume{00}

\firstpage{1}

\journalname{Nuclear Physics A}

\runauth{}

\jid{nupha}

\jnltitlelogo{Nuclear Physics A}

\usepackage{amssymb}

\usepackage[figuresright]{rotating}

\begin{document}

\begin{frontmatter}



\dochead{XXVIIth International Conference on Ultrarelativistic Nucleus-Nucleus Collisions\\ (Quark Matter 2018)}

\title{Chiral Magnetic Effect in Isobaric Collisions from Anomalous-Viscous Fluid Dynamics (AVFD)}


\author[add1]{Shuzhe Shi} 
\author[add2,add1]{Hui Zhang} 
\author[add2]{Defu Hou}
\author[add1,add2]{Jinfeng Liao$^{\, * \,}$\corref{cor1}} 

\address[add1]{Physics Department and Center for Exploration of Energy and Matter,
Indiana University, 2401 N Milo B. Sampson Lane, Bloomington, IN 47408, USA.}
\address[add2]{Institute of Particle Physics and Key Laboratory of Quark \& Lepton Physics (MOE), Central China Normal University, Wuhan, 430079, China.}

 \cortext[cor1]{Presenter}

\begin{abstract} 
The isobaric collision experiment at RHIC provides the unique opportunity to detect the possible signal of Chiral Magnetic Effect (CME) in heavy ion collisions. The  idea is to contrast the correlation observables of the two colliding systems that supposedly have identical flow-driven background contributions while quite different CME signal contributions due to the 10\% variation in their  nuclear charge and thus magnetic field strength. With the recently developed quantitative simulation tool for computing CME signal, the Anomalous-Viscous Fluid Dynamics (AVFD), we demonstrate that a joint (multiplicity + elliptic-flow) event selection is crucial for this purpose. We further propose to use the absolute difference  between RuRu and ZrZr events (after using identical event selection) for detecting CME signal and make predictions for the correlation observables.  
\end{abstract}


\end{frontmatter}


\section{Introduction}
\label{}

The Chiral Magnetic Effect (CME) is a novel quantum transport phenomenon in chiral materials, predicting the generation of an electric current along an external magnetic field under the presence of chirality imbalance. The CME has been observed in the so-called Dirac and Weyl semimetals and has been  studied across a broad range of other physical systems. Enthusiastic efforts have also been made to search for the CME in the quark-gluon plasma (QGP) created by relativistic heavy ion collisions at the Relativistic Heavy Ion Collider (RHIC) and the Large Hadron Collider (LHC)~\cite{STAR_CME,STAR_CME_BES,ALICE_CME,CMS_CME}. An unambiguous observation of CME would provide the direct experimental evidence for the restoration of chiral symmetry in the hot QGP and for the gluon topological fluctuations  in Quantum Chromodynamics (QCD).  For  reviews see e.g. \cite{Kharzeev:2015znc,Bzdak:2012ia}. 

In the context of heavy ion collisions, the CME induces an electric current along the magnetic field arising mainly from the fast-moving spectator protons. The azimuthal orientation of this field is approximately perpendicular to the reaction plane (RP)~\cite{Bloczynski:2012en}, therefore the CME current will lead to a charge separation across the reaction plane. This signal can be measured by the  azimuthal correlations of same-sign (SS) and opposite-sign (OS) charged hadron pairs. The observable is defined as $\gamma^{\alpha\beta}=\langle \cos \left( \phi^\alpha + \phi^\beta - 2\Psi_{RP} \right) \rangle$, which is sensitive to the CME contributions. However it also suffers from various background contributions. In fact an analysis of the $\gamma$ correlator together with another observable $\delta^{\alpha\beta}=\langle \cos \left( \phi^\alpha - \phi^\beta \right) \rangle$, reveals that the measured $\gamma$-correlator is dominated by backgrounds~\cite{Bzdak:2009fc}. 
The extraction of CME signal from $\gamma$-correlator has been extremely challenging.  

To address this issue, quantitative modeling of CME signal is badly needed. To describe CME in heavy ion collisions, one needs a new hydrodynamic framework that accounts for the anomalous transport currents from CME.   Recently a significant step forward has been achieved  through the development of the  Anomalous-Viscous Fluid Dynamics (AVFD) framework~\cite{Shi:2017cpu,Jiang:2016wve,Yin:2015fca},  which is  the state-of-the-art simulation tool for the quantitative computation of CME signal. Using the AVFD, many important features of the CME signal have been characterized, see detailed results in \cite{Shi:2017cpu}. 

Furthermore it is important to develop new experimental approaches that can help decipher signal from backgrounds~\cite{QM18,Magdy:2017yje,Magdy:2018lwk,Wen:2016zic,Zhao:2017nfq,Xu:2017qfs,Skokov:2016yrj}. In particular the R-correlator method, proposed by Magdy, Lacey, et al~\cite{Magdy:2017yje,Magdy:2018lwk},  shows the promising property of significantly suppressing the backgrounds in the observable. 

The isobaric collision experiment~\cite{Skokov:2016yrj}, recently completed in the 2018 run at RHIC, provides the unique opportunity to detect the possible signal of Chiral Magnetic Effect (CME) in heavy ion collisions. The idea is to contrast the correlation observables of the two colliding systems that supposedly have identical flow-driven background contributions while quite different CME signal contributions due to the 10\% variation in their  nuclear charge and thus magnetic field strength. In this contribution, we 
use the AVFD to   demonstrate that a joint (multiplicity + elliptic-flow) event selection is crucial for a successful isobaric experimental analysis.  We further propose to use the absolute difference  between RuRu and ZrZr events (after using identical event selection) for detecting CME signal and make predictions for  the correlation observables.


\section{Event Selection for the Isobaric Collisions: Insights from Initial Conditions}

The key for success of the isobaric contrast idea, is to make sure that {\em one has two  collections of events from RuRu and ZrZr  respectively  that must be identical in their bulk properties (in particular the multiplicity and elliptic flow $v_2$)}. Conventionally one would select events based on centrality and then compare RuRu with ZrZr at same centrality. It is however not clear whether that would be sufficient, particularly given the uncertainty in the nuclear geometry of the isobars~\cite{Deng:2016knn,Sun:2018idn} as well as the subtlety of the CME signal.  

\begin{figure}[!hbt] 
\begin{center}
\includegraphics[scale=0.26]{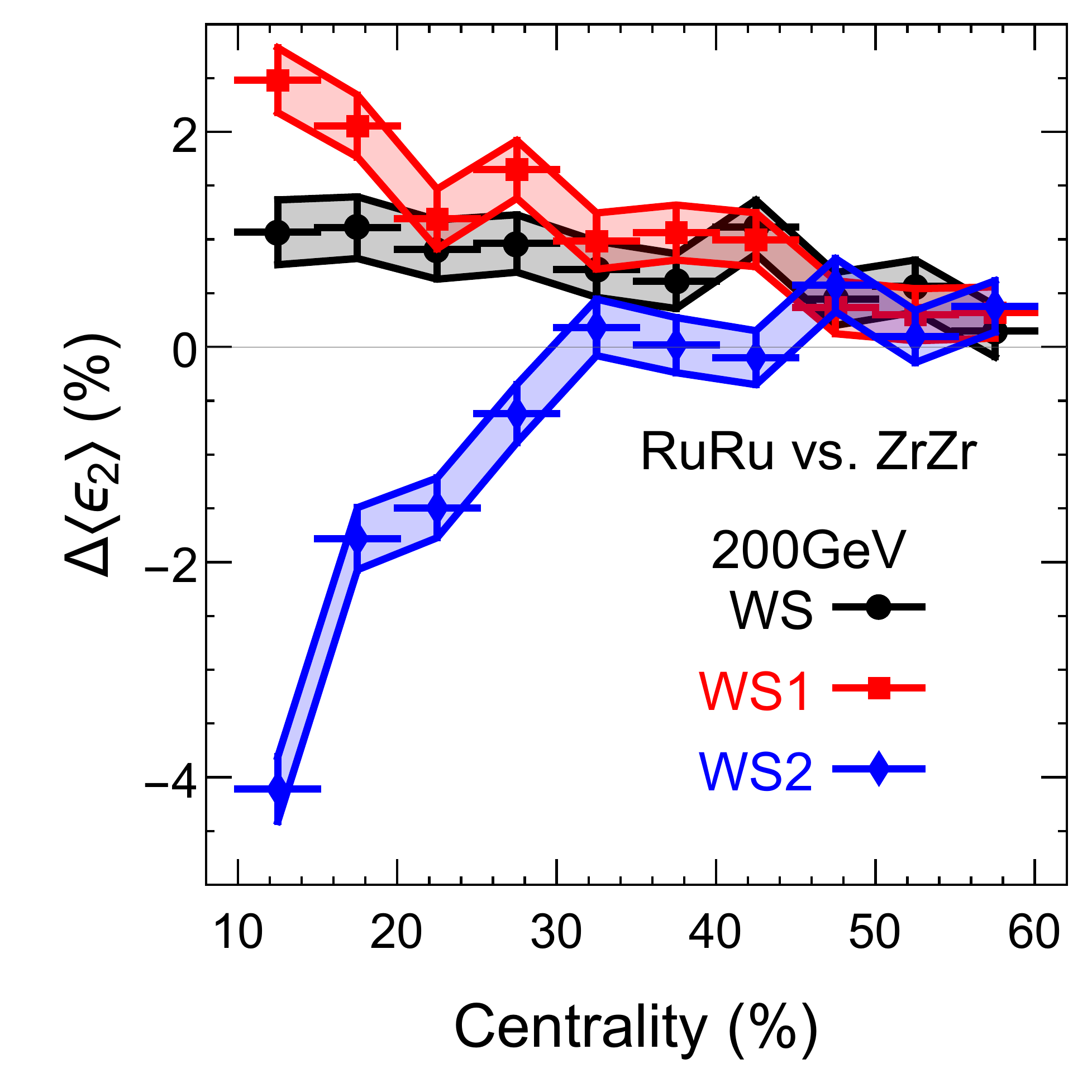}  \hspace{0.3in}
\includegraphics[scale=0.26]{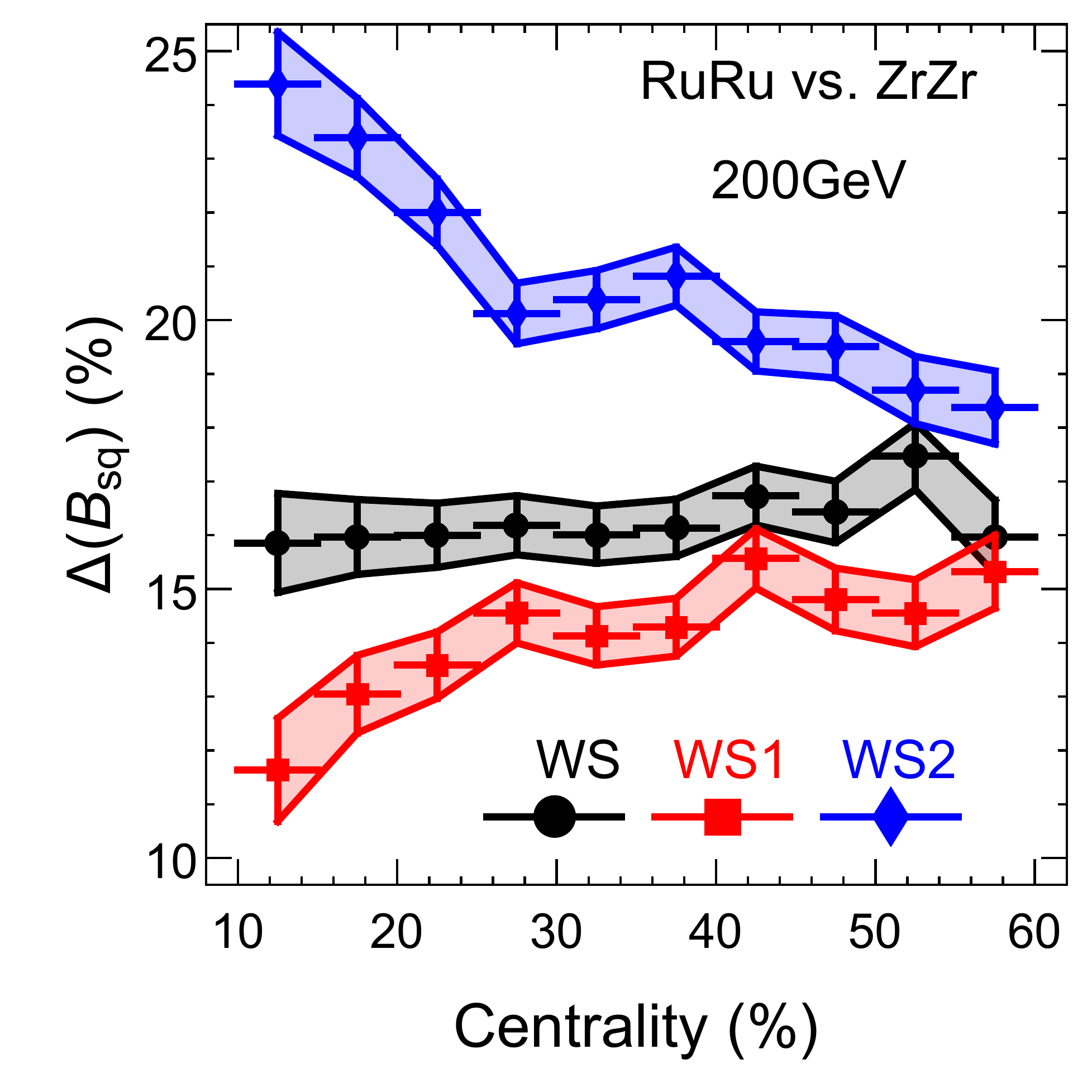}
\caption
{ (Color online) The relative difference  in eccentricity $\Delta\langle \epsilon_2 \rangle$ (left) and   projected magnetic-field-strength-squared $\Delta(B_{sq})$ (right) between RuRu and ZrZr,  with conventional centrality event selection.} 
 \label{fig1}
\vspace{-0.15in}
\end{center}
\end{figure}

To ensure a successful isobar contrast, we propose to  use a joint (multiplicity + elliptic-flow) event selection method.   Let us first use the initial conditions to demonstrate the importance of such selection. In Fig.~\ref{fig1} we show the comparison between the initial conditions of RuRu versus ZrZr with usual centrality class event selection: left panel for relative difference in eccentricity $\Delta\langle \epsilon_2 \rangle$ while right panel for relative difference in projected magnetic-field-strength-squared $\Delta(B_{sq})$. The simulations are performed with Monte-Carlo Glauber model, with three different Woods-Saxon (WS) nuclear distributions, with WS for the spherical case, the WS1 for the case of more deformation in Ru nucleus while the WS2 for the case of ore deformation in Zr nucleus~\cite{Skokov:2016yrj}.  As one can see from Fig.~\ref{fig1}, within a given centrality class and depending nuclear geometry, the two systems could still have a few percent level of difference in their elliptic flow which is the main cause of background contributions in $\gamma$-correlator. Their relative difference in magnetic field strength, which drives the signal, is at the level of $10\sim25\%$. The ``separation'' between $\Delta\langle \epsilon_2 \rangle$  and $\Delta(B_{sq})$ is about one order of magnitude, which may or may not be enough. This critically depends on the signal-to-background ratio in $\gamma$: if the signal fraction is very small, then the idea of contrasting the isobar pairs may not work well. 
 
\begin{figure}[!hbt] 
\begin{center}
\includegraphics[scale=0.28]{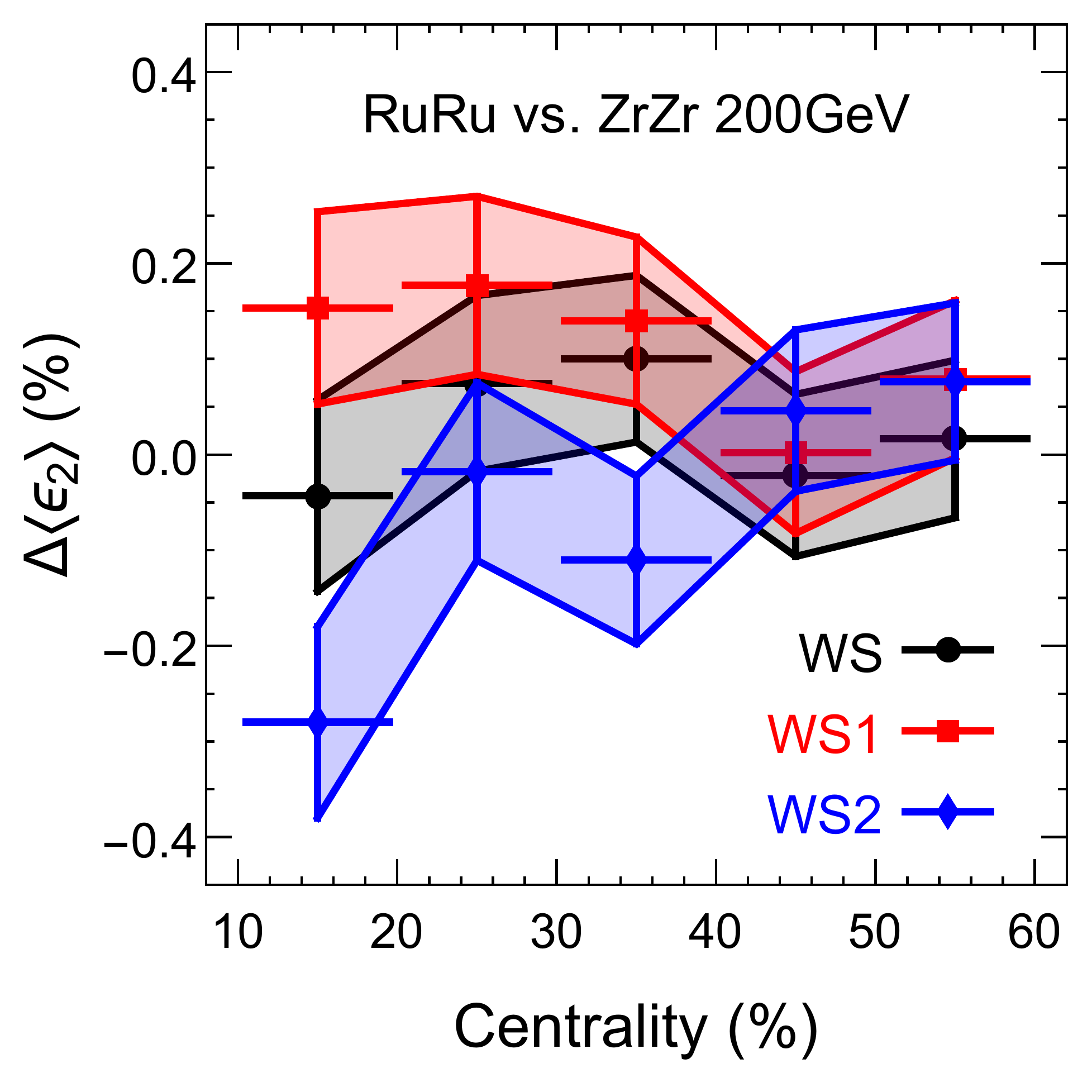}  \hspace{0.3in}
\includegraphics[scale=0.28]{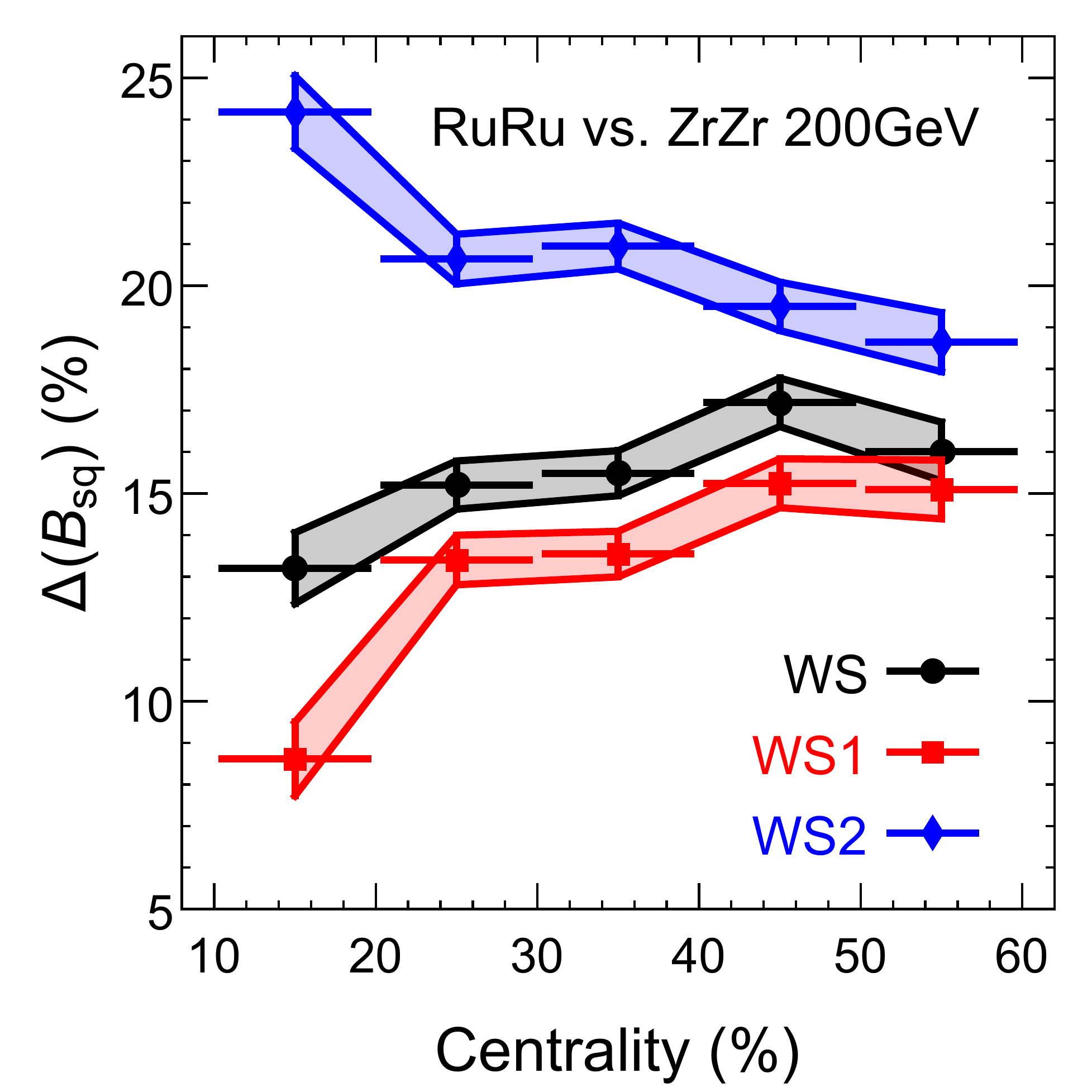}
\caption
{ (Color online) The relative difference  in eccentricity $\Delta\langle \epsilon_2 \rangle$ (left) and   projected magnetic-field-strength-squared $\Delta(B_{sq})$ (right) between RuRu and ZrZr,  with the proposed  joint (multiplicity + elliptic-flow) event selection.  } 
 \label{fig2}
\vspace{-0.25in}
\end{center}
\end{figure}

Therefore, to address this issue by comparing two truly identical collections of isobar events, we propose to  use a {\em joint (multiplicity + elliptic-flow) event selection} method. This can be done by (a) first binning all events according to centrality (i.e. multiplicity),  (b) then apply a joint multiplicity-$v_2$ cut for the events within a given centrality. Note both the multiplicity and $v_2$ cut range should be chosen to be around the mean values of that centrality with a span not too wide, so as to select the   ``most typical'' events for a given centrality and drop out the ``weird events'' from unusual fluctuations. The important point is to apply {\em the same cut for both RuRu and ZrZr events}. As a first test of this idea, we use the initial condition samples above and apply such cut on the initial entropy (as proxy for multiplicity) and eccentricity (as proxy for $v_2$). With the so-selected event samples, we compare again the relative difference between RuRu and ZrZr: see Fig.~\ref{fig2}. As one can see: the relative difference in geometry  $\Delta\langle \epsilon_2 \rangle$ is now reduced to the level of $\sim 0.1\%$ while that in magnetic field $\Delta(B_{sq})$ remains at the $10\sim25\%$ level. The separation between the two has now become about two orders of magnitude and  the contrast between RuRu and ZrZr are most likely able to reveal potential magnetic-field-driven CME signal.

\section{Predictions for CME Signals in Isobaric Collisions with EBE-AVFD}

\begin{figure}[!hbt] 
\begin{center}
\includegraphics[width=0.35\textwidth]{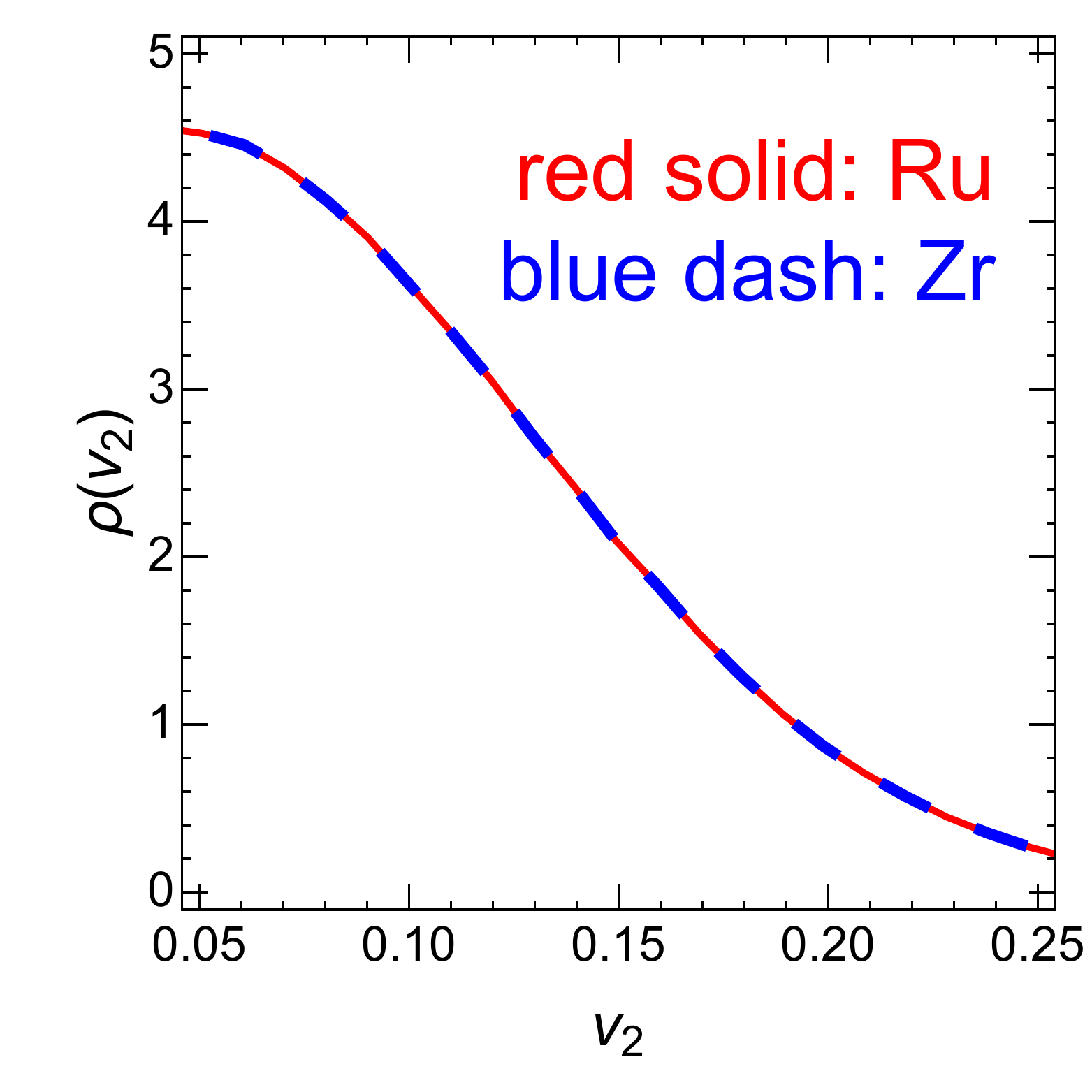}  \hspace{0.3in}
\includegraphics[width=0.35\textwidth]{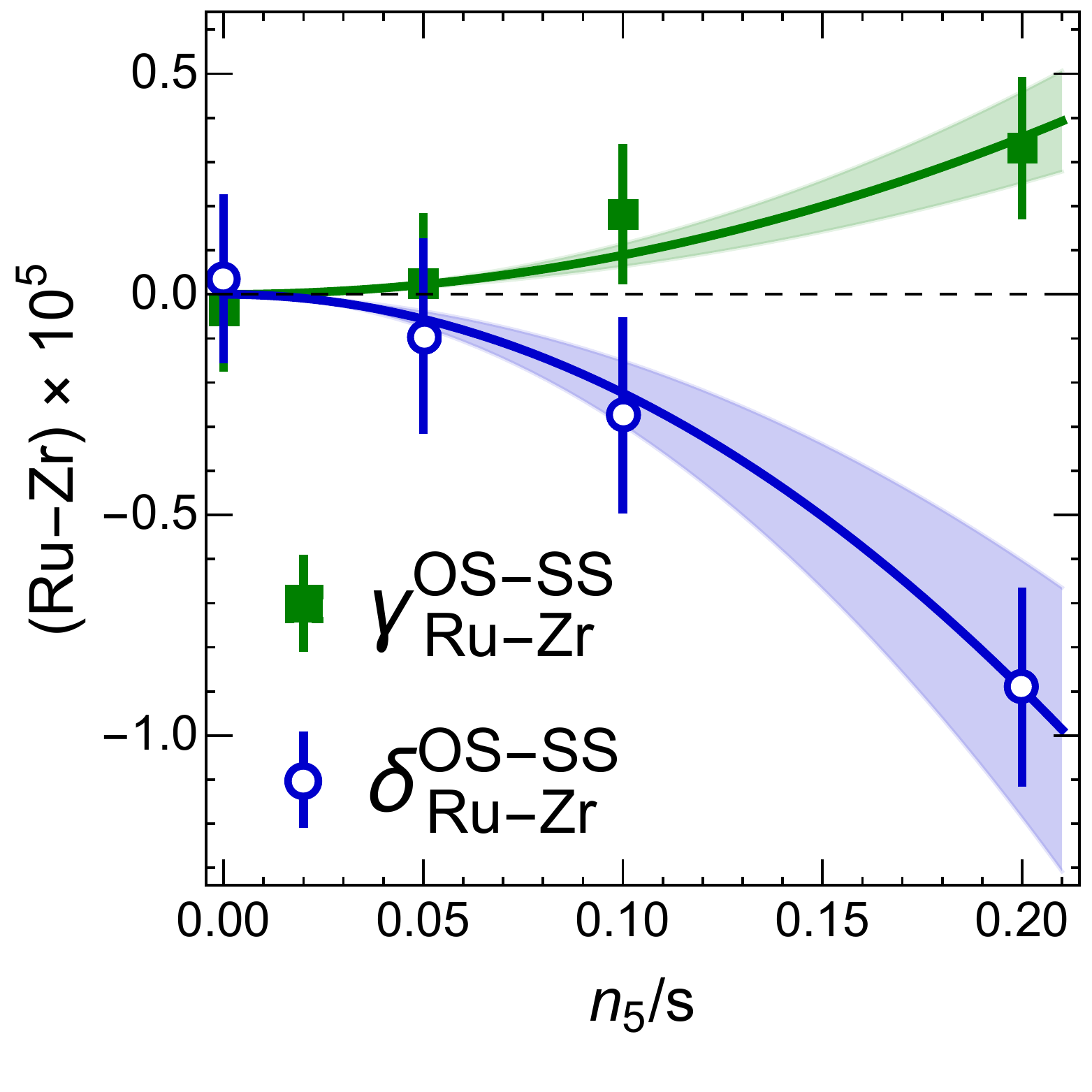}
\caption
{(Color online) (left) The measured $v_2$ distributions of the post-selection simulation events for RuRu and ZrZr collisions. (right) The absolute difference in correlation observables, $(\gamma_{Ru}^{OS-SS}-\gamma_{Zr}^{OS-SS})$ and $(\delta_{Ru}^{OS-SS}-\delta_{Zr}^{OS-SS})$ measured with post-selection events, for varied signal strength as controlled by initial axial charge density $n_5/s$.  } 
 \label{fig3}
\vspace{-0.3in}
\end{center}
\end{figure}

The above discussions focus on insights from initial conditions. To really test the idea and to make predictions for observables, one needs to simulate isobaric collision events and analyze the final state hadrons. To do this, we have upgraded the AVFD into an event-by-event tool (EBE-AVFD)~\cite{EBEAVFD}. For this contribution, we focused on the $40\sim50\%$ centrality and simulated many millions of collision events for both RuRu and ZrZr systems. We then apply the same joint cut: $64< N^{ch}_{|y|<1} <96$ and $0.05 < v_2 < 0.25$ for both systems. With the selected events we then perform physical analysis of correlation observables $\gamma$ and $\delta$. As a consistency check, we compare the measured $v_2$ distributions of the post-selection  events for RuRu and ZrZr, shown in   Fig.~\ref{fig3} (left), which are basically identical and thus guarantee the same background contributions in the isobar pairs. We note  that the $v_2$ cut has the added benefit of  improving event-plane resolution. 


With these post-selection isobar events, we then propose to examine the absolute difference in correlation observables between RuRu and ZrZr~\cite{EBEAVFD}. This is different from   taking a ratio of the $\gamma_{Ru}/\gamma_{Zr}$ , the result of which strongly depends on the signal-to-background ratio within the $\gamma$. Instead, taking the difference would simply subtract out the background portion (which is guaranteed to be identical by the proposed event-selection). The results for $(\gamma_{Ru}^{OS-SS}-\gamma_{Zr}^{OS-SS})$ and $(\delta_{Ru}^{OS-SS}-\delta_{Zr}^{OS-SS})$ are shown in Fig.~\ref{fig3} (right), for varied strength of CME signal which in turn is quadratically dependent on initial axial charge density $n_5/s$. Projections based on present analysis of AuAu collisions~\cite{QM18} would hint at a range $(n_5/s)\sim (5-10)\%$, and a reading from our plot would indicate a measurable absolute difference not only in  $\gamma$-correlator   but also in  $\delta$-correlator which are both sensitive to the presence of CME contributions.  The number of collected isobar events~\cite{QM18} are about 100 times more than that from our AVFD simulations, thus the experimental analysis would have a statistic error bar about 10 times smaller than that shown in this plot. The prospect of being able to ``bracket''   the level of CME signal from isobaric collisions is therefore very encouraging.


\vspace{0.05in}

{\small {\bf Acknowledgments.} 
This material is based upon work supported by the U.S. Department of Energy, Office of Science, Office of Nuclear Physics, within the framework of the Beam Energy Scan Theory (BEST) Topical Collaboration. JL is grateful to the Indiana University Institute for Advanced Study for generous support. The reported work is also supported in part by the NSF Grant No. PHY-1352368 and the NSFC Grant No. 11735007. } 





\vspace{-0.1in}

\bibliographystyle{elsarticle-num}
\bibliography{<your-bib-database>}



\end{document}